\documentclass[12pt,a4paper,british]{iopart}
\usepackage{amsfonts}
\usepackage{amssymb}
\usepackage{color}
\usepackage[T1]{fontenc}
\usepackage[latin9]{inputenc}
\usepackage{babel}
\usepackage{graphicx}
\usepackage{epstopdf}
\usepackage[numbers,square,sort&compress]{natbib}
\usepackage[unicode=true,pdfusetitle,bookmarks=false,breaklinks=false,pdfborder={0 0 1},backref=false,colorlinks=false]{hyperref}
\usepackage{array}
\usepackage{verbatim}
\usepackage{amstext}
\usepackage{iopams}
\usepackage{setstack}

\makeatletter

\newcommand{\bra}[1]{\langle #1|}
\newcommand{\ket}[1]{|#1\rangle}
\newcommand{\braket}[2]{\langle #1|#2\rangle}
\renewcommand{\t}[1]{\textrm{#1}}

\newcommand{\eqref}[1]{(\ref{#1})}

\begin{document}

\title[Super-additivity in communication from a quantum parameter estimation perspective]{Super-additivity in communication of classical information through quantum channels from a quantum parameter estimation perspective}

\author{Jan Czajkowski, Marcin Jarzyna, Rafa{\l} Demkowicz-Dobrza{\'n}ski}
\address{Faculty of Physics, University of Warsaw, ul. Pasteura 5, PL-02-093 Warszawa, Poland}
\eads{\mailto{Marcin.Jarzyna@fuw.edu.pl}, \mailto{demko@fuw.edu.pl}}

\begin{abstract}
We point out a contrasting role the entanglement plays in communication and estimation scenarios. In the first case it
brings noticeable benefits at the measurement stage (output super-additivity), whereas in the latter it is the entanglement
of the input probes that enables significant performance enhancement (input super-additivity). We identify a weak estimation  regime  where a strong connection between concepts crucial to the two fields is demonstrated; the accessible information and the Holevo quantity on one side and the quantum Fisher information related quantities on the other. This allows us to
shed  new light on the problem of super-additivity in communication using the concepts of quantum estimation theory.
\end{abstract}


\maketitle

\section{Introduction}
The greatest achievement of  classical communication theory is realization of the fact that being able to use a noisy communication
channel many times allows one to encode, transmit and decode a message in an error-free way at a non-zero asymptotic rate referred to as the capacity of the channel. A classical channel is described via a conditional probability distribution relating input and output symbols from which the capacity of the channel can be directly calculated \cite{shannon48, cover2012elements}. In a quantum setting \cite{nielsen2010quantum, wilde2013}, two additional elements,
that have an impact on the amount of classical information that can be transmitted through the channel, need to be considered.
The first one is the family of quantum states  that are used to send the encoded information and
the second is the measurement  that provides the read-out. Only then the corresponding conditional probability
can be evaluated and the capacity can be calculated using the classical formula.

\begin{figure}[t]
\begin{center}
\includegraphics[width=0.75 \columnwidth]{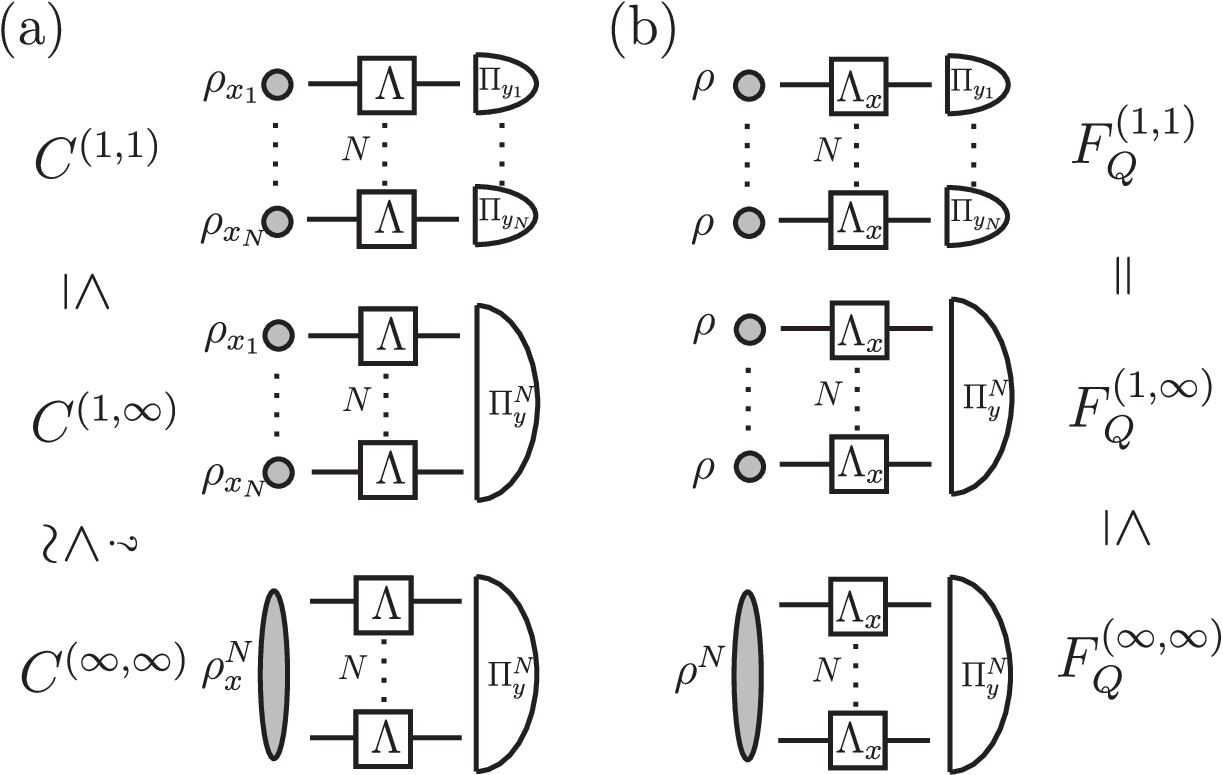}
\end{center}
\caption{General schemes illustrating the role of entangled inputs (input super-additivity) and collective measurements (output super-additivity) in communication (a) and estimation (b) protocols. The performance is quantified by channel capacity $C$ for communication and the quantum Fisher information $F_Q$ for estimation problems,  while the labels in the superscripts inform whether
entanglement is utilized ($\infty$) or not ($1$) at the input and output stages respectively.
While in communication scenarios it is the measurement stage where the super-additivity appears naturally, when the practical role of input super-additivity is debatable, it is the entanglement at the input that offers a significant precision enhancement in case of parameter estimation. Note that an estimation scenario can be regarded as a special case of a communication task where parameter encoding is fixed by the channel parameter dependence $\Lambda_x$.}
\label{fig:scheme}
\end{figure}

There is more to it, however. In a quantum scenario we can imagine
 states entering inputs corresponding to different uses of a channel to be entangled. This may in principle lead to an advantage in communication capacity compared with
a strategy where only separable states are allowed, see figure~\ref{fig:scheme}.
This potential gain thanks to entanglement of input states is referred to as super-additivity of quantum channel capacity and
its actual existence is a topic of long and hot debate in quantum communication community \cite{Schumacher1997, Holevo1998, shor2003, fukuda2007, brandao2007, hastings2009superadditivity, Shor2009, Brandao2010}.
In this paper we will refer to this concept as the \emph{input super-additivity}.
Moreover, even if we do not employ entangled state at the input we are still left with the possibility to perform collective measurements at the output---measure states arriving at different channel outputs coherently.
This may and indeed in many cases does provide a benefit in the form of increased capacity and we will refer to this
effect as the \emph{output super-additivity} \cite{sasaki1998quantum, Yuen2001, Guha2011, Banaszek2012}.

Quantum parameter estimation theory has been largely developed before the quantum communication field achieved its maturity.
These two fields share a common element, they both care to find the measurement optimal for the purpose of extracting classical information
encoded in quantum states. In a communication problem the quantum channel is given, but the character of the information and
the way that it is being encoded in the quantum states is arbitrary and it is ideally chosen in a way to maximize the final information transfer.
In an estimation problem, on the other hand, the information is encoded in quantum states in a particular way
either directly or via the action of some parameter dependent channels, see figure~\ref{fig:scheme}. In this sense the estimation problem may formally be regarded as a restricted communication problem, even though the traditional figures of merit used in estimation and communication approaches are usually different \cite{Yuen2001}.  The central problem of quantum communication and estimation theories is to identify the potential benefits
coming from exploiting entanglement in the input states  as well as at the measurement stage of the protocols.
Interestingly, unlike in the communication problem, there is a great number of examples demonstrating significant gains coming from the
use of entangled input probes in quantum estimation protocols, with applications ranging from optical and atomic interferometry,
via magnetometry to spectroscopy and atomic clocks stabilization \cite{Leibfried2004, Giovannetti2006, Giovannetti2011, demkowicz2015quantum, Paris2009, DAriano2001, Teklu2009}. At the same time, in a typical estimation protocol utilizing unentangled input probes
collective measurements are typically irrelevant.
Thus a contrasting picture emerges:  when thinking of capacities of quantum channels  gains in information processing arising from utilizing entanglement are at the  measurement stage whereas it is the input stage where entanglement makes a difference in the estimation scenarios.

The goal of this paper is to better understand the connections between the two fields from the point of view of the super-additivity issue, understood here as a general question of utility of entanglement in estimation/communication protocols.
We show that in the weak estimation regime, where the amount of information
extractable on the parameter is very small compared to the prior information, i.e. the error of estimation is large compared with the variance of prior distribution,
the accessible information as well as the Holevo quantity can be expressed using Fisher information-like concepts, which allows us to discuss utility of entanglement in communication
using the properties of quantities well understood within the field of quantum estimation.
 We also point out that in a communication problem encoding a large number of independent parameters is favored even at the cost of their limited estimation precision, which makes the estimation strategy of learning a given parameter with highest possible accuracy not likely to be useful for the communication purposes.
In order to make the paper self-contained we also recall some of the recent results on applications of rate-distortion theory in
quantum estimation, which is another way of connecting the communication and estimation fields \cite{nair2012fundamental, hall2012does}.
Let us point out here, however, that while the rate-distortion theory allowed to draw interesting conclusion on performance of certain estimation
protocols using results from communication field our direction of reasoning in this paper is mostly the opposite.
Using results \emph{from} estimation theory we aim at getting some interesting intuitions regarding the communication tasks.
This approach has to be taken with care, since, as pointed out before, the estimation problem is a kind of restricted
communication problem where parameter encoding is already fixed and typically not optimal for communication purposes.
Hence, if certain statements are made on the communication problem they only apply to this particular kind of
information encoding employed in the considered estimation protocol. We therefore cannot claim, and indeed we do not, that our observations
have general implications on the fundamental task of quantum communication theory that is of finding the ultimate capacity of the channel under optimal encoding.
Our main purpose is to elucidate connections between the fields and only if a particular example of encoding considered happens to be optimal
for communication purposes, a connection between fundamental quantum channel capacity and the estimation related quantities can be established, as will be demonstrated in a particular case of communication through bosonic channels.

The paper is organized as follows. Sec.~\ref{sec:communication} and  Sec.~\ref{sec:estimation}
contain a review of relevant concepts from quantum communication and estimation theories highlighting
the role of entanglement in both fields. Sec.~\ref{sec:rdt} discusses known results relating communication and estimation
aspects of quantum information processing via concepts of rate-distortion theory.
Sec.\ref{sec:weak} discusses the weak estimation regime where in \eqref{eq:Mutual_FI_clas} and \eqref{eq:holevofinal} the
 connections between the mutual information and the Holevo quantity respectively with Fisher information-like quantities are established. Sec.\ref{sec:superadditivity} contains discussion of the super-additivity issue from the perspective of this connection. In particular for the weak estimation regime in \eqref{eq:gammasimple} we quantify output superadditivity and then, with the help of advanced tools from estimation theory, we discuss the issue of input superadditivity. In this section we additionally  consider also an opposite regime of strong estimation in which we conjecture \eqref{eq:strongestimationholevo} which imply a lack of output superadditivity and a possible presence of the input one.
 Finally, Sec.\ref{sec:examples} contains examples illustrating the applicability of the weak-estimation approximation and
 discusses its implications on communication via qubit channels in presence of dephasing and bosonic channel under loss and thermal noise.

\section{Communication}
\label{sec:communication}
The main goal of classical communication theory is understanding the limits of sending
credible information through noisy channels. For this purpose the sender needs to appropriately encode the message
and the receiver needs to decode it in a way that the message is not corrupted by the noise of the channel.
 Mathematically, a classical communication channel is modeled by a probabilistic map
 connecting input ($X$) and output ($Y$) random variables via  conditional probability distribution $p(y|x)$, $x \in \mathcal{X} $, $y \in \mathcal{Y}$. According to the Channel Coding Theorem \cite{shannon48, cover2012elements},
  the maximal number of bits that can be correctly transmitted per channel use, referred to as the capacity of the channel $C$,
  reads:
  \begin{equation}
  \label{eq:capacity}
  C=\max_{\{p(x)\}_{x\in \mathcal{X}}}I(X:Y),
 \end{equation}
 where $I(X:Y) = H(Y) - H(Y|X)$ is the Shannon mutual information, $H(Y)\equiv-\sum_{y} p(y)\log p(y)$ is the Shannon entropy of the output and
 $H(Y|X)= -\sum_{x, y} p(y|x)p(x) \log p(y|x)$ is the Shannon conditional entropy.
In this paper all logarithms are assumed to be in base $2$.
It is important to stress that even though
the symbols sent through different independent channels may be correlated, the formula for the capacity is given in terms
of relation between input and output variables of a single channel. This automatically implies that
the capacity $C_N$ of a channel constructed by grouping $N$ individual independent channels into a single entity,
fulfills the additivity property $C_N = N C$.

Quantum communication \citep{holevo2012quantum, wilde2013} is concerned  with sending messages encoded in quantum states through quantum channels, see figure~\ref{fig:scheme}(a). In this paper we will only consider the problem of sending classical messages through quantum channels, ignoring the problem of transmitting faithfully quantum states themselves,
as only this aspect of communication can be expected to have some relation with the estimation problem which in the end deals with extraction of classical parameter encoded in quantum states. Mathematically, a quantum channel is a completely positive trace preserving (CPTP) map $\Lambda$  \cite{nielsen2010quantum} acting on quantum states represented as density operators.
Communication performance of the channel crucially depends on the states $\{\rho_x^{\t{in}}\}_{x \in \mathcal{X}}$ in which we encode input symbols $x$ as well as the operators $\{\Pi_y\}_{y \in  \mathcal{Y}}$ representing the final measurement. To keep full generality one
typically allows for general positive operator valued  measurements (POVM) \cite{nielsen2010quantum}, so that the only condition on
the measurement operators are: $\Pi_y \geq 0$, $ \sum_{y \in \mathcal{Y}} \Pi_y = 1\!\!1$.
With the family of input states as well as the measurement operators fixed, the conditional probability distribution
relating input and output symbols reads $p(y|x) =\Tr(\rho_x \Pi_y)$, where $\rho_x = \Lambda(\rho_x^{\t{in}})$ represent the input states
after they have been transmitted through channel $\Lambda$.
Using now the classical formula for channel capacity, \eqref{eq:capacity}, we get the corresponding
formula for capacity of a quantum channel:
\begin{equation}
\label{eq:capacity11}
C^{(1,1)} = \max_{\{p(x),
\rho_x^{\t{in}},
\Pi_y\}_{x\in \mathcal{X},y\in \mathcal{Y}}}I(X:Y),
\end{equation}
where superscript $^{(1,1)}$ indicates that no entanglement is involved neither at the input nor at the output stage of the protocol.
Unlike in a classical scenario the issue of additivity of the capacity of the quantum channel is far from obvious.
We may both perform collective measurements involving multiple output states as well as send states which
are entangled throughout different channel inputs. This makes classical additivity arguments in general invalid as
the full conditional probability relating input and output symbols of multiple channels no longer factorizes into single channel
quantities.

Indeed, when collective measurements are allowed, the capacity is in general larger than the one given in \eqref{eq:capacity11}
\cite{sasaki1998quantum, Yuen2001, Guha2011, Banaszek2012, pirandola2011quantum}
and is expressed via the so called Holevo quantity $\chi$:
\begin{equation}
\label{eq:capacity1infty}
C^{(1,\infty)} = \max_{\{p(x),
\rho_x^{\t{in}}\}_{x\in \mathcal{X}}} \chi(\{p_x,\Lambda(\rho_x^{\t{in}})\}),
\end{equation}
where
\begin{equation}
\label{eq:holevo}
\chi(\{p_x,\rho_x\}) = S\left(\sum_x p(x) \rho_x\right)- \sum_{x} p(x) S(\rho_x),
 \end{equation}
with $S(\rho) = - \Tr(\rho \log \rho)$ being the von Neuman entropy. The replacement of $1$ with $\infty$ in the right superscript represents the possibility of measuring collectively arbitrary number of
output channels. Apart from covering a more general scenario the above formula has also a clear advantage over \eqref{eq:capacity11} as it no longer requires optimization over measurements.

When the input states are additionally allowed to be entangled, one can also formally write a formula for the capacity
using regularization of the Holevo quantity \citep{Schumacher1997}
\begin{equation}
C^{(\infty,\infty)} = \lim_{N \rightarrow \infty} \frac{1}{N} \max_{\{p(x),
\rho_x^{N,\t{in}}\}_{x\in \mathcal{X}^{\times N}}} \chi(\{p_x,\Lambda^{\otimes N}(\rho_x^{N, \t{in}})\}),
\end{equation}
which is, however, infeasible to deal with due to the necessity of considering entangled quantum states of
arbitrary large number of subsystems.
In case of commonly encountered quantum channels it is proven or at least strongly expected based on numerical investigations that $C^{(\infty,\infty)} =  C^{(1,\infty)}$ \cite{Hayashi2005, Datta2005, Wolf2005}.
The overall picture is more complicated, however, due to the example of Hastings \cite{hastings2009superadditivity} where
a construction of two channels is given for which the Holevo quantity is demonstrated to be strictly super-additive.
The construction is probabilistic and deals with channels of potentially very high dimensions, and
as a result it is hard to assess the quantitative impact of thus demonstrated super-additivity for practical communication scenarios.
To the best knowledge of the authors, up till know there has been no explicit example of a low dimensional channel relevant for communication purposes for which input super-additivity would be demonstrated. Therefore in this paper, we will
write that $C^{(\infty,\infty)} \overset{?}{\gtrsim} C^{(1,\infty)}$ which is supposed to represent the fact that while
there are input super-additive properties of the Holevo quantity, up till now they have not been demonstrated
to be relevant in  practical communication scenarios.

To summarize this section we may therefore write the following relation:
\begin{equation}
\label{eq:comsummary}
C^{(\infty,\infty)} \overset{?}{\gtrsim} C^{(1,\infty)} \geq C^{(1,1)},
\end{equation}
pointing out to the presence of the output and the practical lack of the input super-additivity when thinking of the classical capacity of quantum channels.

\section{Estimation}
\label{sec:estimation}
Classical estimation theory provides methods to optimally estimate a value of a parameter $x$ based on observations $y$
that are known to be distributed according to probability distribution $p(y|x)$, that
represents the probabilistic model for the problem considered. For this purpose
one looks for the optimal estimator function $\tilde{x}(y)$ that minimizes the estimated parameter deviation
from the true parameter value.
Identifying the optimal estimator is non-trivial and
its form in general critically depends on the  prior knowledge available.
Nevertheless, assuming the estimator is unbiased:
$\langle \tilde{x} \rangle = \sum_y p(y|x)\tilde{x}(y) = x $---so that it on average returns the true value---the Cram{\'e}r-Rao (CR) inequality \citep{kay1993fundamentals} allows to write a lower bound on the estimator variance:
\begin{equation}
\label{eq:cr}
\Delta^2 \tilde{x}  \geq \frac{1}{F(x)}, \qquad F(x)  = \sum_y \frac{\dot{p}(y|x)^2}{p(y|x)},
\end{equation}
where $\Delta^2 \tilde{x} = \sum_y p(y|x)(\tilde{x}(y)-x)^2$, dot denotes differentiation with respect to $x$ and $F$ is the Fisher information (FI). Provided one identifies an estimator saturating the above bound one is sure to have found the optimal
one. Even though saturation of the bound is possible for only a very limited class of probability functions, the so called exponential family of distributions \cite{kay1993fundamentals, lehmann2003theory}, the situation is much clearer in the asymptotic regime when one registers many
observations $y_i$, $i \in \{1,\dots,N\}$ independently and identically distributed according to $p(y_i|x)$.
In this case the joined probability distribution $p(y_1,\dots,y_N|x) = \Pi_{i=1}^N p(y_i|x)$ is product
and hence, thanks to additivity of FI on independent probability distributions, the corresponding
FI for $p(y_1,\dots,y_N|x)$ equals $F_N = N F$. As a result
the estimation variance based on $N$ observation is bounded according to \eqref{eq:cr} as:
\begin{equation}
\Delta^2 \tilde{x}_N \geq \frac{1}{N F(x)}.
\end{equation}
Most importantly, the above bound is saturable in the asymptotic limit of $N \rightarrow \infty$ and
the optimal estimator is the max-likelihood estimator \cite{kay1993fundamentals, lehmann2003theory} .
Saturability of the CR bound for large $N$ is intimately related with
the local asymptotic normality theorem \cite{van2000asymptotic} proving that, in the limit of large $N$
and after a suitable reparametrization, probability distribution $\Pi_{i=1}^N p(y_i|x)$
can be viewed as a Gaussian distribution with mean being shifted by $\sqrt{N} x$ and the variance equal to $1/F$.
Since Gaussian distribution with its mean as a parameter to be estimated is a member of the exponential
family of distributions for which the CR bound is saturated this proves the fact.

In a typical \emph{quantum} estimation problem we are given a family of states $\{\rho_x\}$ with the task of learning the parameter $x$.
Apart from the issue of finding the optimal estimator $\tilde{x}(y)$ we also need to find the optimal measurement
$\{\Pi_y\}$ that yields the actual conditional probability distribution of observed results $p(y|x) = \Tr(\rho_x \Pi_y)$.
The quantum generalization of the CR inequality yields the lower bound on the achievable variance irrespectively of the measurement
applied \citep{holevo2011probabilistic}:
\begin{equation}
\Delta^2 \tilde{x} \geq \frac{1}{F_Q(\rho_x)}, \qquad F_Q (\rho_x) = \Tr \rho_{x} L_x^2, \label{eq:fqsingdef}
\end{equation}
where $F_Q$ is the quantum Fisher information (QFI) and $L_x$ is the symmetric logarithmic derivative (SLD) operator implicitly defined by:
\begin{equation}
\dot{\rho}_{x} = \frac{1}{2} \left( \rho_{x} L_x + L_x \rho_{x} \right) \label{eq:sldsingdef}.
\end{equation}
When written explicitly in terms of eigenvalues and eigenvectors of $\rho_x = \sum_{n=1}^d p_n \ket{n}\bra{n}$, the QFI reads:
\begin{equation}
\label{eq:qfiexplicit}
F_Q(\rho_x) =  \sum_{n=1}^d\frac{\dot{p}_n^2}{p_n}+2\sum_{n,m=1\atop p_n+p_m \neq 0}^d \frac{(p_n-p_m)^2}{p_n+p_m}|\braket{n}{\dot{m}}|^2,
\end{equation}
where both the eigenvalues and eigenvectors in general depend on $x$ and in case of pure states estimation, $\rho_x = \ket{\psi_x}\bra{\psi_x}$,
the above formula simplifies to
\begin{equation}
\label{eq:qfipure}
F_Q(\ket{\psi_x}) = 4 \left( \braket{\dot{\psi_x}}{\dot{\psi_x}} - |\braket{\dot{\psi_x}}{\psi_x}|^2 \right).
\end{equation}
QFI is additive on product states, so that $F(\rho_x^{\otimes N}) = N F(\rho_x)$. Hence,
given $N$ copies of the state we get:
\begin{equation}
\Delta^2 \tilde{x}_N \geq \frac{1}{N F_Q(\rho_x)}.
\end{equation}
Moreover, when a measurement is chosen so that it is a projective measurement in the SLD eigenbasis, the corresponding
FI equals the QFI. Therefore, applying the arguments from from the classical case, the above quantum CR inequality is also asymptotically saturable in the limit of large $N$.

If the family of states to be considered is not given, but the parameter to be estimated is rather encoded in the action of a
channel $\Lambda_x$, we are additionally challenged to find
the optimal probe states $\rho^{\t{in}}$ that allow the parameter $x$ to be estimated with smallest possible uncertainty by measuring the output states $\rho_x = \Lambda_x(\rho^{\t{in}})$. When the probe state is sent into inputs of $N$ copies of the channel $\Lambda_x$,
 one can again ask whether entangled input states and collective measurements
offer any advantage compared to uncorrelated strategies. The answer to this question is the key to understanding the benefits of quantum enhanced estimation scenarios, see figure~\ref{fig:scheme}(b).

If only product input states are allowed, the maximal QFI per channel use reads:
\begin{equation}
F_Q^{(1,1)} = F_Q^{(1,\infty)}  = \max_{\rho^{\t{in}}} F_Q[\Lambda_x(\rho^{\t{in}})],
\end{equation}
where the equality $F_Q^{(1,1)} = F_Q^{(1,\infty)}$ arises thanks to additivity of QFI on product states and the fact that
there always exist a local measurement for which FI equals to the corresponding QFI \citep{braunstein1994statistical}. Therefore, unlike
in the communication case, there is no benefit from application of collective measurements when product input states
are used. However, when inspecting QFI for entangled input probes there are a plethora of examples when entanglement
at the input increases the resulting QFI. In particular, for unitary parameter estimation the QFI may increase at a rate proportional to $N^2$  rather than linearly in $N$ \cite{Giovannetti2006, Giovannetti2011, demkowicz2015quantum} and even though decoherence typically
reduces the asymptotic scaling again to a linear one, the entanglement enhancement benefit remains in terms of a larger multiplicative constant \cite{Escher2011, demkowicz2012elusive, demkowicz2014using}. Hence, in general  QFI is input super-additive and we can therefore write:
\begin{equation}
F_Q^{(\infty,\infty)} \geq F_Q^{(1,\infty)} =  F_Q^{(1,1)},
\end{equation}
where $F_Q^{(\infty,\infty)}$ denotes QFI optimized over all entangled states at the input, which contrasts the analogous relation for communication capacities given in \eqref{eq:comsummary}. It is important to keep in mind, however, that for entangled input states, the optimal detection strategy in some instances may be collective \cite{micadei2015coherent}. Note that we
 take the convention where $F_Q^{(\infty,\infty)}$ denotes the QFI per channel use to make it more like the capacity concept introduced before. The issue of finding $F^{(\infty,\infty)}$ has been addressed in \cite{Hayashi2011, Escher2011, demkowicz2012elusive, Kolodynski2013}.

Up till now we have based our whole discussion of the quantum estimation problem on the analysis of the QFI.
When communication and estimation approaches are to be related, however, it is more natural to adopt the Bayesian perspective on estimation, as the prior distributions of the parameters to be estimated naturally translate to input symbol probability distributions  in a communication problem. Taking the quadratic cost as a figure of merit, the optimal Bayesian estimation of a parameter $x$, given
the familiy of states $\rho_x$ distributed according to the prior $p(x)$,
 is the one that minimizes the \emph{average} variance:
\begin{equation}
\overline{\Delta^2 \tilde{x}}= \int \t{d}x\,  p(x) \int \t{d}y p(y|x)(\tilde{x}(y)-x)^2
\end{equation}
over the choice of measurement operators $\{\Pi_y\}$ and estimators $\tilde{x}(y)$.
A general solution to the problem is known and the minimal achievable variance
equals to \cite{helstrom1976quantum, macieszczak2014bayesian}:
\begin{equation}
\label{eq:BayesianGaussian}
\overline{\Delta^2 \tilde{x}}= \Delta^2_0  - \Tr(\bar{\rho} L^2),
\end{equation}
where  $\bar{x} = \int \t{d}x\, p(x) x$ is the mean and $\Delta^2_0 = \int \t{d}x \ p(x)(x - \bar{x})^2$ the variance of the prior
whereas $L$ is implicitly defined via the following relation:
\begin{equation}
\label{eq:sldBayesian}
\bar{\rho}^\prime = \frac{1}{2}(\bar{\rho}L + L \bar{\rho}),
\end{equation}
with $\bar{\rho}$, $\bar{\rho}^\prime$ defined as
\begin{equation}
 \bar{\rho} = \int \t{d}x\, p(x) \rho_x,   \ \bar{\rho}^\prime = \int \t{d}x\, p(x) x \rho_x.
\end{equation}
The apparent similarity of \eqref{eq:sldBayesian} to the SLD formula \eqref{eq:sldsingdef} becomes
even stronger when the prior $p(x)$ is assumed to be Gaussian in which case
\eqref{eq:BayesianGaussian} becomes:
\begin{equation}
\label{eq:BayesianGaussianFinal}
\overline{\Delta^2 \tilde{x}} =  \Delta^2_0 \left[1 - \Delta^2_0 F_Q(\bar{\rho})\right],
\end{equation}
with $F_Q(\bar{\rho})$ being the QFI for the problem of estimating the mean of the prior $\bar{x}$ given
the averaged state $\bar{\rho}$.

The Bayesian perspective is indeed adopted in papers making use of rate-distortion theory to derive
bounds on estimation precision using communication tools \cite{nair2012fundamental, hall2012does}---see section \ref{sec:rdt}.
Fortunately, the conclusions on the role of entanglement in the quantum estimation problem discussed above using the QFI concept remain qualitatively unchanged when the Bayesian methodology is applied \cite{demkowicz2015quantum}. Hence it is often enough to study the properties of the QFI which is easier to analyze. The QFI related quantities will also prove useful in Sec.\ref{sec:weak} where
it is demonstrated that they play an important role in analyzing communication performance in the weak estimation regime.

\section{Rate-distortion theory}
\label{sec:rdt}
A first natural place to look for relations between estimation and communication problems is the rate-distortion theory \cite{shannon1959coding, cover2012elements}.
The main objective of the rate-distortion theory is to quantify how much information can be transmitted provided given level of errors and vice versa. In particular, viewing the estimation protocol as a communication channel form the
 input symbol $x$ to its estimator $\tilde{x}$, it is possible to lower bound the
 corresponding mutual information $I(X:\tilde{X})$ via \cite{cover2012elementsA}
\begin{equation}
\label{eq:rtdbound}
I(X:\tilde{X}) \geq H(X) -\frac{1}{2} \log(2 \pi e \overline{\Delta^2 \tilde{x}}),
\end{equation}
where $\overline{\Delta^2 \tilde{x}}$ is the average estimation variance and for continuous random variables $H(X)$ denotes a differential entropy.
Intuitively, this relation reflects the fact that the better the estimation precision the higher the communication rate.
Or stated the other way round, one needs to communicate a lot in order to estimate very precisely.
Note, that here we refer to the estimation problem using Bayesian perspective as we explicitly take into account
 the form of prior distribution. Recall also that in the whole paper we focus on transmitting classical information encoded in quantum systems and therefore utilize results of classical rate-distortion theory abstracting from a more general quantum rate-distortion theory \cite{datta2013quantum} where faithful communication of quantum states themselves is considered.

Utilizing \eqref{eq:rtdbound} together with the fact that $I(X:\tilde{X})$ is upper bounded by the Holevo quantity $\chi$
one can get a lower bound on the achievable estimation variance \cite{nair2012fundamental, hall2012does, hall2013metrology}
\begin{equation}
\label{eq:rdtBayes}
\overline{\Delta^2 \tilde{x}} \geq \frac{4^{H(X)- \chi(\{p_x,\rho_x\})}}{2\pi e}.
\end{equation}
Thinking of states $\rho_x$ as the outputs of a parameter dependent channel $\rho_x = \Lambda_x(\rho^{\t{in}})$,
we may obtain the lower bound $\overline{\Delta^2 \tilde{x}}$ valid for arbitrary input probe states, provided
we are able to upper bound the corresponding Holevo quantity $\chi[\{p_x,\Lambda_x(\rho^{\t{in}})\}]$.
A good candidate is the capacity $C^{(1,\infty)}$, \eqref{eq:capacity11}, of the channel
which is obviously an upper bound for $\chi$. The problem is that typically the formula for the capacity is not easily obtained
 and also the resulting bound may not be very informative.
For example, for a single mode lossy bosonic channel with effective transmission $\eta$ it is known
that if the average number of photons at the input is upper bounded by $\bar{n}$ the capacity of the channel reads:
\begin{equation}
C^{(1,\infty)} = C^{(\infty,\infty)}=(\eta \bar{n} +1)\log(\eta \bar{n} +1) - \eta \bar{n} \log(\eta \bar{n}).
\end{equation}
When plugged into \eqref{eq:rdtBayes} this yields in the large $\bar{n} \gg 1$ limit:
$
\overline{\Delta^2 \tilde{x}} \geq \frac{4^{H(X)}}{2 \pi e^3} \frac{1}{(\eta \bar{n})^2}.
$
Thinking now of phase estimation, the bound is reasonably tight for the lossless case $\eta=1$. However, in case
of losses ($\eta <1$), the bound is highly unsatisfactory since it is known that the Heisenberg limit is lost \cite{demkowicz2015quantum}, and the achievable asymptotic scaling of phase estimation variance is $1/\bar{n}$ rather than $1/\bar{n}^2$. This is related to the fact that the optimal encoding that saturates the capacity
of the channel is not the phase encoding characteristic for the phase estimation problem.
Therefore, instead of plugging in the capacity of the channel itself it may be more reasonable to insert a tighter bound on $\chi$ obtained
for an encoding present in a given estimation problem.
Following this way of reasoning, a much more informative bound has been derived for the case of unitary parameter estimation $\Lambda_x(\rho^{\t{in}})=U_x \rho^{\t{in}} U_x^\dagger$, $U_x = \exp(-\mathrm{i} G x)$, \cite{hall2012does}:
\begin{equation}
\overline{\Delta^2 \tilde{x}} \geq \frac{1}{2\pi e} 4^{H(X)+S(\rho^{\t{in}}) - H(G|\rho^{\t{in}})},
\end{equation}
where $H(G|\rho^{\t{in}})$ is the Shannon entropy of the measurement statistics corresponding to measuring
$\rho^{\t{in}}$ in the eigenbasis of the generator $G$. This approach allowed to obtain useful precision bounds
in case of decoherence-free nonlinear quantum metrology \citep{hall2012does}
 and lossy optical estimation \cite{nair2012fundamental}, where the correct $1/\bar{n}$ phase estimation variance scaling in presence of losses has been recovered.

The results summarized above made used of connections between estimation and communication fields in order
to obtain original results in estimation theory.  In this paper we
focus on a complementary goal. We aim to obtain a better insight into communication aspects of quantum channels
 benefiting from our understanding of estimation related quantities.

\section{Weak estimation regime}
\label{sec:weak}
In this section we identify a regime where a connection between Shannon and Holevo quantities on one side and the QFI related quantities on the other can be established. This regime corresponds to a situation which we refer to as \emph{the weak estimation regime}.
The precise conditions will be given further in this section, but intuitively the regime we are interested in
corresponds to a situation in which the knowledge on the parameter gained from measurement of the output state is small compared to the prior knowledge. More formally, we can state this condition as an assumption that $\Delta_0^2\ll1/F(\bar{x})$, where $\Delta_0^2$ is a variance of the prior distribution and $F(\bar{x})$ is Fisher information of the conditional probability distribution $p(y|x)$ defining the channel evaluated at the prior mean value $\bar{x}$. This regime is indeed of physical interest in some important instances of communication, especially communication on large distances when the power of the incoming signal is weak; we analyze a particular example of such case in Sec.\ref{sec:examples}. Importantly, note that weak estimation regime usually do not apply to a situation in which we send the same symbol many times since then we can learn a lot about the parameter. This is reflected in the fact that Fisher information of the total conditional distribution increases proportionally to the number of channel uses which leads to breaking the condition $\Delta^2_0\ll1/F(\bar{x})$. We specifically consider this opposite regime in Sec.\ref{sec:strong_est_regime}.

Our main results relate mutual information and Holevo quantity with Fisher information and its quantum counterparts. In the classical case we show that in the weak estimation regime \cite{Abbe2010, Huang2015}
\begin{equation}
I(X:Y)\approx \frac{\Delta_0^2}{2 \ln 2} F(\bar{x}),
\end{equation}
whereas in the quantum case we have
\begin{equation}
\chi \approx  \frac{\Delta_0^2}{2 \ln 2} \underline{J}(\rho_{\bar{x}}) - \sum_{n=r+1}^d \frac{\Delta^2_0 F_n(\rho_{\bar{x}})}{4}
\log \frac{\Delta^2_0 F_n (\rho_{\bar{x}})}{4 e},
\end{equation}
where $\underline{J}$ is a quantity analogical to QFI but with slightly different operational meaning and $F_n$ can be directly related to QFI. The exact definition of $\underline{J}$ and $F_n$, proofs of the above equations and discussion will be given further in this section.

\subsection{Classical case}
Let us first discuss the classical case and write mutual information between the sender and the receiver
\begin{eqnarray}\label{eq:mutinf_D}
\nonumber I(X:Y) =H(Y) - H(X|Y) = \\
- \int  \t{d}y \, p(y)\log p(y) + \int \t{d}x \t{d}y \,p(x)p(y|x)\log p(y|x).
\end{eqnarray}
Assuming $p(y|x)$ is sufficiently smooth in $x$ and the prior $p(x)$ is sufficiently narrow we approximate
$p(y|x)$ using expansion around the prior mean $\bar{x}$ up to the second order $p(y|x) \approx p(y|\bar{x})+\dot{p}(y|\bar{x})(x-\bar{x})+\frac{1}{2}\ddot{p}(y|\bar{x})(x-\bar{x})^2$, where dots denote derivatives
with respect to $x$ taken at $x=\bar{x}$. Taking the expectation value of this expression with respect to the prior $p(x)$
we obtain $p(y)\approx p(y|\bar{x})+\frac{\Delta^2_0 \ddot{p}(y|\bar{x})}{2}$, where $\Delta^2_0$ is the prior variance.
Using this approximation, the first term in  \eqref{eq:mutinf_D} hence reads:
%
\begin{equation}
-\int \t{d}y\,
\left(p(y|\bar{x})+\frac{\Delta^2_0 \ddot{p}(y|\bar{x})}{2}\right)\log \left(p(y|\bar{x})+\frac{\Delta^2_0 \ddot{p}(y|\bar{x})}{2}
\right).
\end{equation}
We now expand the $\log$ function and keep the leading order terms in $\Delta^2_0$ arriving at:
\begin{equation}
\label{eq:secondterm}
H(Y) \approx
 -\int \t{d}y\,\left[\left( p(y|\bar{x}) + \frac{\Delta^2_0 \ddot{p}(y|\bar{x})}{2}\right)\log p(y|\bar{x}) +
 \frac{\Delta^2_0 \ddot{p}(y|\bar{x})}{2 \ln 2}\right].
\end{equation}
By doing so, we have made an implicit assumption that $p(y|\bar{x}) > 0$, otherwise the expansion would not be possible.
Within our order of approximation, ignoring contribution from terms for which $p(y|\bar{x})=0$, is justified provided in this cases
$\dot{p}(y|\bar{x})=\ddot{p}(y|\bar{x})=0$ as well. This is the technical assumption which intuitively means that
events which are impossible when $x=\bar{x}$ do not gain probability too rapidly when moving away from $\bar{x}$.

Moving on to the second term. We expand the conditional entropy around $\bar{x}$ up to the second order in $(x-\bar{x})$:
\begin{eqnarray}\fl
\nonumber\int \t{d}y \,p(y|x)\log p(y|x) \approx \int \t{d} y\,  \bigg[p(y|\bar{x})\log p(y|\bar{x}) + \\
\nonumber\left(\dot{p}(y|\bar{x}) \log p(y|\bar{x}) + \frac{\dot{p}(y|\bar{x})}{\ln 2}\right)(x-\bar{x}) + \\
\frac{1}{2}\left(\ddot{p}(y|\bar{x}) \log p(y|\bar{x}) + \frac{\dot{p}(y|\bar{x})^2}{p(y|\bar{x}) \ln 2} + \frac{\ddot{p}(y|\bar{x})}{\ln 2} \right) (x-\bar{x})^2\bigg].
\end{eqnarray}
Taking now the average of the above expression over the prior $p(x)$, the linear term vanishes and the result reads:
\begin{eqnarray}
\label{eq:firstterm}
\nonumber H(Y|X) \approx -\int \t{d} y\,  \bigg[p(y|\bar{x})\log p(y|\bar{x}) + \\
-\frac{\Delta^2_0}{2}\left(\ddot{p}(y|\bar{x}) \log p(y|\bar{x}) + \frac{\dot{p}(y|\bar{x})^2}{p(y|\bar{x}) \ln 2} + \frac{\ddot{p}(y|\bar{x})}{\ln 2} \right)\bigg].
\end{eqnarray}
Subtracting  \eqref{eq:firstterm} from \eqref{eq:secondterm} we arrive at:
\begin{equation}\label{eq:Mutual_FI_clas}
I(X:Y)\approx \frac{\Delta_0^2}{2 \ln 2} F(\bar{x}),
\end{equation}
where $F(\bar{x})$ is the FI of $p(y|x)$ evaluated at $\bar{x}$.

Note, that in the above derivation, while expanding the logarithm in the expression for Shannon entropy $H(Y)$ we have assumed that $\frac{\Delta^2_0}{2\ln 2}\frac{\ddot{p}(y|x)|_{\bar{x}}}{p(y|\bar{x})} \ll 1$. In order to expand logarithm in the expression for conditional entropy $H(Y|X)$ we additionally assumed also that $\Delta^2_0 \ll 1/F(\bar{x})$
meaning the prior variance is much smaller than
 the variance dictated by the CR bound. This intuitively means that for the approximation \eqref{eq:Mutual_FI_clas} to hold
 our gain of knowledge on the parameter obtained from the observed data must be small compared to  prior knowledge.

\eqref{eq:Mutual_FI_clas} reminds of a known relation between the FI and the relative entropy.
Relative entropy $D(p{\parallel}q) = \sum_y p(y) \log[p(y)/q(y)]$ is a natural measure of a difference between two probability distributions.
When considering two neighboring probability distributions $p(y|x)$, $p(y|x+\t{d}x)$ their relative entropy is approximated
by the FI up to the second order in $\t{d}x$ \cite{gourieroux1995statistics}:
\begin{equation}
\label{eq:relativeentropyexpansion}
D[p(y|x+\t{d}x)||p(y|x)] \approx  \frac{1}{2 \ln 2}F(x) \t{d}x^2.
\end{equation}
On the other hand mutual information may be expressed via relative entropy as:
\begin{equation}
\label{eq:mutualrelative}
I(X:Y) = \int \t{d} x\, p(x) D[p(y|x) || p(y)].
\end{equation}
Had we replaced $p(y)$ in the above formula with $p(y|\bar{x})$ and expanded relative entropy around $\bar{x}$ using \eqref{eq:relativeentropyexpansion} we would indeed get \eqref{eq:Mutual_FI_clas}. Validity of this replacement
hinges upon the assumption that knowledge of the input parameter to be equal to the prior mean does not alter the conditional probability substantially. This again intuitively corresponds to the weak estimation regime,
but is hard to justify formally without a more detailed analysis as presented above.

\subsection{Quantum case}
Moving now on to the quantum world, we ask for the generalization of \eqref{eq:Mutual_FI_clas}
that would provide a connection between quantum communication and quantum estimation concepts.
The most natural step would be to replace the FI appearing in \eqref{eq:Mutual_FI_clas}
with the QFI. Indeed, this is a right approach provided we use product input states and no collective measurements,
hence the $(1,1)$ scenario. In this case we simply replace single channel probabilities
$p(y|x)$ with $\Tr(\rho_x \Pi_y)$. Since there always exist a measurement for which the corresponding FI equals to the
QFI, this implies that when communicating using $\rho_x$ states with variance of prior distribution much narrower than
$1/F_Q(\rho_{\bar{x}})$ the mutual information may be approximated as:
\begin{equation}
\label{eq:informationfisherindividual}
\max_{\{ \Pi_y\}} I(X:Y)  \approx \frac{\Delta_0^2}{2 \ln 2}F_Q(\rho_{\bar{x}}).
\end{equation}
As a side remark, note that utilizing inequality \eqref{eq:rtdbound}, assuming a Gaussian prior and making use
of an explicit relation between the the Bayesian cost and the QFI given in \eqref{eq:BayesianGaussianFinal} leads to:
\begin{equation}
\max_{\{ \Pi_y\}}  I(X:Y) \geq H(X) - \frac{1}{2} \log[2 \pi e \Delta^2_0(1-\Delta^2_0 F_Q(\bar{\rho})].
\end{equation}
Since for Gaussian prior $H(X) = \frac{1}{2}\log(2 \pi e \Delta^2_0)$, we get:
\begin{equation}
\max_{\{ \Pi_y\}}  I(X:Y)\geq  - \frac{1}{2} \log[1-\Delta^2_0 F_Q(\bar{\rho})] \approx \frac{\Delta^2_0}{2 \ln2} F_Q(\bar{\rho}).
\end{equation}
Where the right hand side of the above inequality differs from \eqref{eq:informationfisherindividual} only by the  replacement of $\rho_{\bar{x}}$ with $\bar{\rho}$. Clearly this makes sense, as mixing a state cannot increase the QFI, and hence the above inequality is indeed in agreement with our approximation.

Let us now consider the Holevo quantity given by \eqref{eq:holevo} describing  communication capabilities when collective measurements are allowed. First note, that the Holevo quantity may be expressed as
\begin{equation}
\label{eq:holevorelativeentropy}
\chi(\{p_x,\rho_x\}) = \int \t{d} x\, p(x) D(\rho_x|| \bar{\rho}),
\end{equation}
where $\bar{\rho}=\int \t{d}x\, \rho_x$ is the average state and
\begin{equation}
D(\rho||\sigma) = \Tr\rho (\log \rho - \log \sigma)
\end{equation}
is the quantum relative entropy \cite{nielsen2010quantum}.
Interestingly, when expanding the quantum relative entropy
for neighboring quantum states up to the second order we get \cite{petz2002covariance}
\begin{equation}
\label{eq:quantumrelativeentropyfisher}
D(\rho_{x+\t{d} x}||\rho_{x})\approx  \frac{1}{2 \ln 2}J(\rho_x)\t{d} x^2
\end{equation}
where
\begin{equation}
\label{eq:reqfi}
J(\rho_x)=\sum_{n=1}^d\frac{\dot{p}_n^2}{p_n}+2\sum_{n,m=1}^d(p_n-p_m)|\braket{n}{\dot{m}}|^2\ln p_n,
\end{equation}
with $p_n$ and $\ket{n}$ being the eigenvalues and eigenvectors of $\rho_x$ and $d$ the dimension of the Hilbert space. Comparing the above equation with \eqref{eq:qfiexplicit} it is clear that $J(\rho_x)$ is in general \emph{not} equal to the QFI,
and we will refer to it as the relative entropy quantum Fisher information (REQFI). In fact it upper bounds the respective QFI
$J(\rho_x)\geq F_Q(\rho_x)$, with equality for diagonal density matrices \cite{petz2002covariance}. Moreover, REQFI gives meaningful results only on mixed states, being infinite on pure states. This last fact is a counterpart of the infiniteness of quantum relative entropy for pure states.

Proceeding by analogy to the classical case, one might attempt to replace  $\bar{\rho}$ with $\rho_{\bar{x}}$ in \eqref{eq:holevorelativeentropy},
plug in the expansion \eqref{eq:quantumrelativeentropyfisher} and arrive at an approximate formula for Holevo quantity
as in  \eqref{eq:informationfisherindividual} but with QFI replaced by the REQFI. Instead, we provide below a general derivation
for the approximating formula for the Holevo quantity, which proves that the above heuristic argument only works in case
of full rank states (or states which are effectively full rank in the sense that their kernel subspace can be trivially removed from the considerations) and is not justified in general. Intuitively,
this is related with the fact that $\bar{\rho}$ which is obtained as a probabilistic mixture
may in general be a state of higher rank than $\rho_{\bar{x}}$, and has a significant impact on the Holevo quantity.

Taking the Holevo quantity
\begin{equation}
\label{eq:Holevo_Rel}
\chi=S(\bar{\rho})- \int dx p(x) S(\rho_x)
\end{equation}
 we  expand $\rho_x$ around the prior mean $\bar{x}$ up to the second order and get $\rho_x\approx \rho_{\bar{x}} +\dot{\rho}_{\bar{x}}(x-\bar{x})+\frac{1}{2}\ddot{\rho}_{\bar{x}}(x-\bar{x})^2$.
 The average state at the output is therefore equal to $\bar{\rho}=\int \t{d}x\, p(x)\rho_x\approx \rho_{\bar{x}}+\frac{\Delta_0^2 \ddot{\rho}_{\bar{x}}}{2}$.
 Let $p_n$ denote eigenvalues and $\ket{n}$ denote eigenbasis of $\rho_{\bar{x}}$, which in case of degeneracy is further
 specialized to diagonalize $\ddot{\rho}_{\bar{x}}$ on each degenerate subspace.
 To calculate the first term in \eqref{eq:Holevo_Rel}
   we only need to know eigenvalues $\bar{p}_n$ of $\bar{\rho}$. Treating $\frac{\Delta_0^2\ddot{\rho}_{\bar{x}}}{2}$ term as a small perturbation added to $\rho_{\bar{x}}$ we make use of standard perturbation theory and get that up to the first-order correction
   $\bar{p}_n\approx p_n+\frac{\Delta_0^2 (\ddot{\rho}_{\bar{x}})_{nn}}{2}$, where $(\ddot{\rho}_{\bar{x}})_{nn} = \bra{n} \ddot{\rho}_{\bar{x}}\ket{n}$ and hence
\begin{equation}
\label{eq:holevofirstterm}
S(\bar{\rho})\approx -\sum_{n=1}^d\left(p_n+\frac{\Delta_0^2 (\ddot{\rho}_{\bar{x}})_{nn}}{2}\right)\log\left(p_n+\frac{\Delta_0^2 (\ddot{\rho}_{\bar{x}})_{nn}}{2}\right).
\end{equation}
We make analogous assumption as in the classical derivation, namely that eigenvalues that are zero when $x=\bar{x}$ do
not grow too rapidly when moving away from $\bar{x}$. Hence we assume that if $p_n =0$ then also $\dot{p}_n=\ddot{p}_n=0$.
Still the situation we face is significantly different than in the classical case.
Even with the assumptions made, we are not  entitled to neglect the terms for which $p_n=0$ since
$\dot{p}_n=\ddot{p}_n=0$ does not imply that $(\ddot{\rho}_{\bar{x}})_{nn}=0$.
This is a crucial point and is related to the intrinsically quantum transformation of the states---the unitary transformation.
Let $r \leq d$ be the rank od $\rho_{\bar{x}}$. We split \eqref{eq:holevofirstterm} into
two parts depending on whether $p_n$ is strictly positive ($1 \leq n \leq r$)  or zero ($r+1 \leq n  \leq d$)
and expand the logarithm whenever it is positive
\begin{eqnarray}
\label{eq:holevofirsterm}
\nonumber S(\bar{\rho})\approx -\sum_{n=1}^r\left(p_n+\frac{\Delta_0^2 (\ddot{\rho}_{\bar{x}})_{nn}}{2}\right)\log p_n +
\frac{\Delta_0^2 (\ddot{\rho}_{\bar{x}})_{nn}}{2 \ln 2}+ \\
-\sum_{n=r+1}^d \frac{\Delta_0^2 (\ddot{\rho}_{\bar{x}})_{nn}}{2} \log \frac{\Delta_0^2 (\ddot{\rho}_{\bar{x}})_{nn}}{2}.
\end{eqnarray}

Moving on to the second term in \eqref{eq:Holevo_Rel}, note that $S(\rho_x)$ only depends on eigenvalues of
$\rho_x$ and its expansion is identical as in the classical case. Hence after averaging over $p(x)$ we get
\begin{eqnarray}
\label{eq:holevosecondterm}
\nonumber\int \t{d}x p(x) S(\rho_x) \approx \\ \approx -\sum_{n=1}^r \bigg[p_n\log p_n + \frac{\Delta^2_0}{2}\bigg(\ddot{p}_n\log p_n+\frac{\dot{p}_n^2}{p_n \ln 2} + \frac{\ddot{p}_n}{\ln 2}\bigg)\bigg],
\end{eqnarray}
where the sum is over non-zero $p_n$.
Subtracting \eqref{eq:holevosecondterm} from \eqref{eq:holevofirstterm} we get:
\begin{eqnarray}
\chi \approx \sum_{n=1}^r \frac{\Delta_0^2}{2 \ln 2}\left([\ddot{p}_n - (\ddot{\rho}_{\bar{x}})_{nn}][\ln p_n+1] + \frac{\dot{p}_n^2}{p_n}\right) + \\
-\sum_{n=r+1}^d \frac{\Delta_0^2 (\ddot{\rho}_{\bar{x}})_{nn}}{2} \log \frac{\Delta_0^2 (\ddot{\rho}_{\bar{x}})_{nn}}{2}.
\end{eqnarray}
Thanks to trace preservation $\sum_{n=1}^r \ddot{p}_n =0$, and $\sum_{n=1}^d (\ddot{\rho})_{nn} =0$ (note the summation upper limit is $d$),
and as a result:
\begin{eqnarray}
\label{eq:holstepalmostfinal}
\nonumber\chi \approx \sum_{n=1}^r \frac{\Delta_0^2}{2 \ln 2}\left([\ddot{p}_n - (\ddot{\rho}_{\bar{x}})_{nn}]\ln p_n + \frac{\dot{p}_n^2}{p_n}\right) + \\
-\sum_{n=r+1}^d \frac{\Delta_0^2 (\ddot{\rho}_{\bar{x}})_{nn}}{2} \left(\frac{1}{\ln 2} + \log \frac{\Delta_0^2 (\ddot{\rho}_{\bar{x}})_{nn}}{2}\right).
\end{eqnarray}
Writing $(\ddot{\rho}_{\bar{x}})_{nn}$ explicitly we have
\begin{equation}
(\ddot{\rho}_{\bar{x}})_{nn} = \ddot{p}_n - 2 p_n \braket{\dot{n}}{\dot{n}} + 2\sum_{k=1}^r p_k |\braket{n}{\dot{k}}|^2.
\end{equation}
Note that $\braket{\dot{n}}{\dot{n}} = \sum_{k=1}^d |\braket{\dot{n}}{k}|^2$
and since $\ket{k}$, $\ket{n}$ denote orthonormal eigenvectors then $\dot{\braket{k}{n}}=0$ and hence we can replace $\braket{\dot{n}}{k}$ with $- \braket{n}{\dot{k}}$ arriving at:
\begin{equation}
(\ddot{\rho}_{\bar{x}})_{nn} = \ddot{p}_n + 2 \sum_{k=1}^d (p_k - p_n)|\braket{n}{\dot{k}}|^2.
\end{equation}
Plugging the above formula into  \eqref{eq:holstepalmostfinal} and recalling the definition of REQFI, \eqref{eq:reqfi}, we arrive at the final expression:
\begin{equation}
\label{eq:holevofinal}
\chi \approx  \frac{\Delta_0^2}{2 \ln 2} \underline{J}(\rho_{\bar{x}}) - \sum_{n=r+1}^d \frac{\Delta^2_0 F_n(\rho_{\bar{x}})}{4}
\log \frac{\Delta^2_0 F_n (\rho_{\bar{x}})}{4 e},
\end{equation}
where the underline symbol in $\underline{J}$ indicates that the sum in the definition of $J$ is restricted to $n \leq r$, avoiding zero $p_n$, and thus, unlike $J$ which may sometimes be infinite, $\underline{J}$ is always finite. The $F_n$ quantities appearing in the second term
read:
\begin{equation}
\label{eq:partialqfi}
F_n(\rho_{\bar{x}}) = 2 (\ddot{\rho}_{\bar{x}})_{nn}.
\end{equation}
In order to interpret them note that when differentiating the definition of SLD, \eqref{eq:sldsingdef}, we get:
\begin{equation}
\ddot{\rho}_{\bar{x}} = \frac{1}{2}\left(\dot{\rho}_{\bar{x}} L_{\bar{x}} + \rho_{\bar{x}} \dot{L}_{\bar{x}} + \dot{L}_{\bar{x}} \rho_{\bar{x}} + L_{\bar{x}} \dot{\rho}_{\bar{x}}\right)
\end{equation}
and using \eqref{eq:sldsingdef} again yields
\begin{equation}
\ddot{\rho}_{\bar{x}} = \frac{1}{2}\left( L_{\bar{x}} \rho_{\bar{x}} L_{\bar{x}}
+ \frac{1}{2}(\rho_{\bar{x}} L_{\bar{x}}^2 + L_{\bar{x}}^2\rho_{\bar{x}}) + \rho_{\bar{x}} \dot{L}_{\bar{x}} + \dot{L}_{\bar{x}} \rho_{\bar{x}}\right).
\end{equation}
Sandwiching with $\ket{n}$,  which is outside the support of $\rho_{\bar{x}}$ ($n \geq r+1$),  and plugging into
\eqref{eq:partialqfi} we arrive at
\begin{equation}
F_n(\rho_{\bar{x}}) = \bra{n} L_{\bar{x}} \rho_{\bar{x}} L_{\bar{x}}  \ket{n}.
\end{equation}
Comparing this with the definition of the QFI, \eqref{eq:fqsingdef}, which can be rewritten as:
\begin{equation}
F_Q(\rho_{\bar{x}}) = \Tr(\rho_{\bar{x}} L_{\bar{x}}^2) = \sum_{n=1}^d \bra{n} L_{\bar{x}} \rho_{\bar{x}} L_{\bar{x}} \ket{n},
\end{equation}
we see that $F_n$ represent contributions to QFI from the subspace laying outside the support of $\rho_{\bar{x}}$---the kernel of $\rho_{\bar{x}}$.
Recall that the eigenbasis $\ket{n}$ outside the support of $\rho_{\bar{x}}$ is not arbitrary but
was assumed to diagonalize $\ddot{\rho}_{\bar{x}}$.

The approximate expression for Holevo quantity, Eq~\eqref{eq:holevofinal}, simplifies in two special cases.
When $\rho_{\bar{x}}$ is full rank, or $\ddot{\rho}$ lives on the support of $\rho_{\bar{x}}$, the second term
in Eq~\eqref{eq:holevofinal} vanishes and the Holevo quantity only depends on $J$:
\begin{equation}
\label{eq:holevomixed}
\chi \approx  \frac{\Delta_0^2}{2 \ln 2} J(\rho_{\bar{x}}).
\end{equation}
Going to the other extreme, if $\rho_{\bar{x}}=\ket{\psi_{\bar{x}}}\bra{\psi_{\bar{x}}}$ is pure then
the SLD can be written explicitly:
\begin{equation}
L_{\bar{x}} = 2 (\ket{\psi_{\bar{x}}} \bra{\dot{\psi}_{\bar{x}}} + \ket{\dot{\psi}_{\bar{x}}} \bra{{\psi}_{\bar{x}}} )
\end{equation}
and so:
\begin{eqnarray}
\nonumber L_{\bar{x}}\rho_{\bar{x}} L_{\bar{x}} = 4 \bigg(\ket{\dot{\psi}_{\bar{x}}} \bra{\dot{\psi}_{\bar{x}}} +
 |\braket{\dot{\psi}_{\bar{x}}}{\psi}|^2 \ket{{\psi}_{\bar{x}}} \bra{{\psi}_{\bar{x}}} + \\
 \braket{\dot{\psi}_{\bar{x}}}{\psi_{\bar{x}}} \ket{\psi_{\bar{x}}}\bra{\dot{\psi}_{\bar{x}}} +
 \braket{{\psi}_{\bar{x}}}{\dot{\psi}_{\bar{x}}} \ket{\dot{\psi}_{\bar{x}}}\bra{{\psi}_{\bar{x}}}\bigg).
\end{eqnarray}
Thanks to $\braket{\dot{\psi}_{\bar{x}}}{\psi_{\bar{x}}} + \braket{\psi_{\bar{x}}}{\dot{\psi}_{\bar{x}}} =0$ identity
we have $\bra{\psi} L \rho L \ket{\psi} =0$ and hence the whole contribution
to QFI comes from the kernel of $\rho_{\bar{x}}$. Let $P_0$ be projector on the kernel of $\rho_{\bar{x}}$, then:
\begin{equation}
2 P_0 \ddot{\rho}_{\bar{x}} P_0  =  P_0 L_{\bar{x}}\rho_{\bar{x}} L_{\bar{x}} P_0 = 4 P_0 \ket{\dot{\psi}_{\bar{x}}} \bra{\dot{\psi}_{\bar{x}}} P_0.
\end{equation}
Let us write $\ket{\dot{\psi}_{\bar{x}}} = a \ket{\psi_{\bar{x}}} + b \ket{\psi^{\perp}_{\bar{x}}}$, where
$\ket{\psi^{\perp}_{\bar{\psi}}}$ is orthogonal to $\ket{\psi_{\bar{x}}}$. It is now clear that the $\ket{\psi^{\perp}_{\bar{x}}}$
is a proper choice of eigenvector in the kernel subspace that makes $\ket{\dot{\psi}_{\bar{x}}}$ diagonal on this subspace.
Moreover, this is the only vector that will yield any contribution to QFI, and hence there is only one non zero $F_n$ ($n=1+1$) which reads:
\begin{equation}
F_{2} = 4 |\braket{{\psi^{\perp}_{\bar{x}}}}{\dot{\psi}_{\bar{x}}}|^2 =
4(\braket{\dot{\psi}_{\bar{x}}}{\dot{\psi}_{\bar{x}}} - |\braket{{\psi_{\bar{x}}}}{\dot{\psi}_{\bar{x}}}|^2) = F_Q
\end{equation}
and is equal to the QFI, see  \eqref{eq:qfipure}.
Summarizing, for pure state protocols Holevo quantity can be approximated using
only the QFI as:
\begin{equation}
\label{eq:holevopure}
\chi \approx -\frac{\Delta^2_0 F_Q(\ket{\psi_{\bar{x}}})}{4}
\log \frac{\Delta^2_0 F_Q(\ket{\psi_{\bar{x}}})}{4 e}.
\end{equation}

\section{Manifestations of super-additivity}\label{sec:superadditivity}

\subsection{Weak estimation regime}

 Approximate formulas for the mutual information, \eqref{eq:informationfisherindividual}, as well as for the Holevo quantity, \eqref{eq:holevofinal}, provide an interesting insight into the issues of super-additivity.
Let us first assume that no entanglement is used at the input, and that the output
states coming out of individual channels are $\rho_x = \Lambda_x(\rho^{\t{in}})$.
 Let $C_{p(x),\Lambda_x}$ denote the ``capacity'' of the channel under a fixed encoding defined by the prior as well as
 channel parameter dependence $\{p(x), \Lambda_x\}$.
 Note that when talking about capacity,  we implicitly consider a scenario where the above specified encoding is repeated independently over $N$ channels.
 By independently, one should understand here that
 the full action of $N$ channels is $\Lambda_{x_1} \otimes \dots  \otimes \Lambda_{x_N}$, where
 $x_i$ are i.i.d. distributed according to the prior distribution considered.
Assuming the prior is narrow enough so that our approximations hold, and
 we restrict ourselves to only individual measurements at the output
 then invoking the approximate formula \eqref{eq:informationfisherindividual} for the mutual information yields:
 \begin{equation}\label{eq:capacity_individual}
 C^{(1,1)}_{p(x),\Lambda_x} \approx  \max_{\rho^{\t{in}}}  \frac{\Delta^2_0 F_Q[\Lambda_{\bar{x}}(\rho^{\t{in}})]}{2 \ln 2}.
 \end{equation}
 If, however, collective measurements are allowed then utilizing \eqref{eq:holevofinal} we get:
 \begin{eqnarray}\label{eq:capacity_collective}
 \nonumber C^{(1,\infty)}_{p(x),\Lambda_x} \approx   \max_{\rho^{\t{in}}} \frac{\Delta_0^2}{2 \ln 2} \underline{J}(\Lambda_{\bar{x}}(\rho^{\t{in}}))+\\ - \sum_{n=r+1}^d \frac{\Delta^2_0 F_n[\Lambda_{\bar{x}}(\rho^{\t{in}})]}{4} \log \frac{\Delta^2_0 F_n [\Lambda_{\bar{x}}(\rho^{\t{in}})]}{4 e},
 \end{eqnarray}
 where the rank $r$ here refers to the rank of $\rho_{\bar{x}} = \Lambda_{\bar{x}}(\rho^{\t{in}})$.
 Comparing the above two formulas
 one can easily appreciate the advantages coming from the use of collective measurements.
 Let us define a natural measure of output super-additivity as the ratio of the two capacities
\begin{equation}\label{eq:gamma_mix}
\gamma^{(1,\infty)} = \frac{C^{(1,\infty)}_{p(x),\Lambda_x}}{C^{(1,1)}_{p(x),\Lambda_x}}.
 \end{equation}
Focusing for clarity on the two extreme cases of full rank and pure output states,
in which simple approximate formulas (\ref{eq:holevomixed},\ref{eq:holevopure}) for Holevo quantity are valid,
the super-additivity measure reads
\begin{equation}
\label{eq:gammasimple}
\gamma^{(1,\infty)} \approx \cases{\frac{J[\Lambda_{\bar{x}}]}{F_Q[\Lambda_{\bar{x}}]} & \textrm{full-rank}\\ -\frac{\ln 2}{2}\log\frac{\Delta^2_0  F_Q[\Lambda_{\bar{x}}]}{4e}  &\textrm{pure}},
\end{equation}
where $F_Q[\Lambda_{\bar{x}}]$ and $J[\Lambda_{\bar{x}}]$ denote maximal QFI and REQFI of the channel $\Lambda_{\bar{x}}$, i.e. $F_Q[\Lambda_{\bar{x}}]=\max_{\rho^{\t{in}}}F_Q[\Lambda_{\bar{x}}(\rho^{\t{in}})]$ and similarly for REQFI.
Inspecting the full-rank case, we see that the measure of output super-additivity
is equal to the ratio of the REQFI $J$  and the QFI $F_Q$. We have already stressed before that in general
$J \geq F_Q$ and now we can fully appreciate that the gap between these two quantities is actually responsible for
the output super-additivity in the weak communication regime.

On the other hand, in case of pure states, $\gamma^{(1,\infty)}$ is determined solely by the QFI and is divergent in the limit $\Delta^2_0 \rightarrow 0$ indicating a more than a constant factor gain in communication potential thanks to the use of collective measurements.

Note that we can also consider output super-additivity even in the case of fixed state $\rho^{\textrm{in}}$, which results in a fixed set of output states $\rho_x$. In such instance  our measure of super-additivity can be defined similarly as in \eqref{eq:gammasimple} but with QFI and REQFI of the specific state $\rho_{\bar{x}}$ rather than the channel $\Lambda_{\bar{x}}$. This is important in practical applications where usually the set of message states $\rho_x$ is specified by the laboratory apparatus.

Approaching now  the input super-additivity issue, let us consider a scenario where $N$ channels are divided into $k$-channel
 subgroups, $\Lambda^{\otimes k}_x$, that accept $k$-partite entangled probes at their inputs, so that the action of all channels is described as
 $\Lambda^{\otimes k}_{x_1} \otimes \dots \otimes \Lambda^{\otimes k}_{x_{N/k}}$.
 Note that all channels within one subgroup encode the same value of the parameter, and this is repeated independently for other subgroups.
 The corresponding ``capacity'' reads:
 \begin{equation}
 \label{eq:capacitysuperadditivity}
 C^{(k,\infty)}_{p(x),\Lambda_x} = \frac{1}{k} \max_{\rho^{k,\t{in}}}\chi(\{p(x),\Lambda^{\otimes k}_{x}(\rho^{k,\t{in}})\}).
 \end{equation}
 Provided we stay in the regime where our weak communication assumption holds,  the Holevo quantity is
 expressible using $F_Q$ and $J$ and the issue of input super-additivity is therefore
 related to the issue of super-additivity of the QFI and the REQFI.

 Super-additivity of QFI, i.e. the property that $F_Q^{(k,\infty)} >  F_Q^{(1,\infty)}$,
 has been already discussed in Sec.~\ref{sec:estimation} and
 is a typical feature of most quantum channels estimation problems with an exception of a very narrow class, e.g. loss estimation, where no entanglement at the input is needed to reach the optimal performance \cite{monras2007optimal}.
 For example, when ideal unitary parameter estimaion is considered then $F_Q^{(k,\infty)}$ grows linearly in $k$
 and when plugged into pure-state approximate formula for Holevo quantity \eqref{eq:holevopure} the ``capacity'' \eqref{eq:capacitysuperadditivity}
 will get a further boost from the input super-additivity.

 Super-additivity of the REQFI is much less studied, but its relevance is clear from our above analysis as noisy channels
 produce output states
 $\rho_{\bar{x}}$  which are likely to be highly mixed and satisfy the conditions which make the simpler formula for Holevo quantity approximation
 \eqref{eq:holevomixed} valid. While the tools for studying the QFI in noisy channels are highly developed and allow
 to draw immediate conclusions on e.g. the maximal gain that can be offered by entangled probes, analogous studies
 have not been pursued in case of REQFI, as this quantity being an upper bound on QFI typically provides
 looser bounds in quantum estimation problems and hence apart from mathematical investigations was rarely appreciated in
 a more practical-oriented studies. We hope that our present work, will boost the interest in the REQFI, as an element providing a connection
 between estimation and communication problems. For the time being, we provide here an example of reasoning
 that shows how super-additivity of REQFI may be analyzed using tools developed in the  quantum estimation field.

 One of the simplest ideas allowing to find the limit on the maximal achievable QFI when entangled probes are used, is the
 idea of classical simulation of the channel \cite{Matsumoto2010, demkowicz2012elusive}. In this approach, one treats quantum channel space as a probability space over which
 classical parameter dependent probability distribution emulates the quantum channel change, and one upper bounds the QFI by a classical FI of this probability distribution. The basic property of the QFI that is used in this derivation is that it does not increase under
 parameter-independent channels and reduces to classical FI for diagonal density matrices. The same properties are, however, also
 enjoyed by the REQFI. One can therefore immediately apply the known upper bounds on QFI derived using classical simulation method to REQFI.
 Now provided, the bound on QFI is asymptotically saturable, this automatically implies that since $J \geq F_Q$
 the bound is saturable for $J$ as well, and in this way one can obtain an asymptotic formula for $J$ optimized
 over entangled input state of large number of probes proving in particular its input super-additivity.
 We present quantitative results obtained with the help of this kind of reasoning in Sec.~\ref{sec:examples}.

 Therefore, it is clear from the discussion above that in the weak estimation regime, we enjoy both aspects of
 quantum super-additivity; first on the level of measurements (output super-additivity) formally reflected by the
 replacement of $F_Q$ by $J$ or $-\ln 2 F_Q \log \Delta^2_0 F_Q/4e$ in the approximate formula for the Holevo quantity; and second on the level of input probes related to the input super-additivity of $F_Q$ and $J$.

 We should note, however, that the requirements of our approximation never allow us to increase $k$ too much, and least of all consider $k \rightarrow \infty$. This is because our weak estimation assumption requires that
 what we learn at the output is small compared to prior knowledge. Clearly, by increasing $k$ we increase the information available about
 the input parameter and hence at some point the approximation needs to break down.
 In the next subsection, for completeness of our presentation we discuss in more detail
 the opposite regime of $k \rightarrow \infty$, which we refer to as the strong estimation regime and contrast it with
 the weak estimation regime discussed so far.

\subsection{Strong estimation regime}\label{sec:strong_est_regime}
We start with a classical scenario. Consider $k$ independent repetitions of an experiment governed by the
same conditional probability distribution $p(y_i|x)$ so that the joined probability distribution of $k$ results $\boldsymbol{y}^k= \{y_1,\dots,y_k\}$ reads: $p(\boldsymbol{y}^k|x) = \Pi_{i=1}^k p(y_i|x)$.
Note the fixed value $x$ for all experiments.
As mentioned already in Sec.~\ref{sec:estimation}, thanks to the local asymptotic normality theorem \cite{van2000asymptotic}
 we know that in the limit of large $k$, the problem can be translated to that of estimating $x$
  from a Gaussian distribution with mean being shifted by  $\sqrt{k} x$  and variance equal to $1/F(x)$.
  Less formally, we can think of this problem as estimating shift $x$ from a Gaussian distribution with variance narrowing as $1/(k F(x))$.
 This allows to obtain an asymptotic formula for the mutual information in this protocol treating it
  as a communication task of transmitting $x$ \cite{clarke1990information, lehmann2003theory}:
 \begin{eqnarray}
 \nonumber I(X:Y^k) = H(X) - H(X|Y^k) \approx \\ - \int \t{d}x \, p(x) \log p(x) - \int \t{d}x\, p(x) \frac{1}{2}\log\left(2 \pi e \frac{1}{k F(x)}\right) + o(1),
 \end{eqnarray}
 where the second term represents the average entropy of a Gaussian distributions with variance $1/(kF(x))$, and the $o(1)$ is a term of
 order $1$ appearing due to the fact that perfect Gaussianity is achieved only asymptotically.
 The above formula can be rewritten in a more appealing form
 \begin{equation}
  \label{eq:InfBar}
 I(X:Y^k)  \approx  \frac{1}{2} \log \frac{k}{2 \pi e} +  \int \t{d}x\, p(x) \log\frac{\sqrt{F(x)}}{p(x)} + o(1),
 \end{equation}
 which clearly shows that the information communicated increases only logarithmical with $k$, and that it is maximised
 for \emph{Jeffreys prior} $p(x)\sim\sqrt{F(x)}$ \cite{lehmann2003theory}. Jeffreys prior is often considered the least informative prior and therefore it is used to represent a complete lack of knowledge about the parameter \cite{mackay2003information}.

 Due to only logarithmic increase in mutual information, the strong estimation regime is not likely to be
 interesting for communication purposes.
 Given $N$ to be a total number of available uses of a channel,
 it is preferable  to keep $k$ relatively small and repeat the communication procedure $N/k$ times using independently distributed input
 parameters $x_i$. While each of the parameters will not be estimated particularly well, the resulting mutual information will
 scale like $N/k$ quickly surpassing the $\log N$ behaviour which we would be left with had we set $k$ to its maximal available value $N$
 in order to perform the most precise parameter estimation possible. In other words, for the purposes of communication it is much better to have an increasing number of independent parameters transmitted under fixed noise, rather than a fixed number of parameters
 transmitted under decreasing noise.

Considering the quantum counterpart of the above scenario with no entanglement used at the input, i.e. the receiver obtains $\rho_x^{\otimes k}$ state, and no collective measurements allowed at the output, we immediately get an analogue of \eqref{eq:InfBar} by substituting $F(x)$ with $F_Q(\rho_x)$. This is because one can always find a local measurement for which $F=F_Q$.

On the other hand, if we allow for collective measurements we no longer enjoy the product structure of the conditional probability distribution, and therefore cannot directly apply the results of classical local normality theory. Fortunately,
quantum generalization of local asymptotic normality \citep{guctua2006local, gill2013asymptotic} states that the estimation of a parameter from a product state $\rho_x^{\otimes k}$ in the limit of large number of copies $k\to\infty$ is equivalent to estimation of a displacement of a particular Gaussian quantum state with a variance in the direction of the shift corresponding to the QFI---there exist CP maps that translate one problem to another under trace norm. Thanks to continuity of the von-Neuman entropy with respect to the trace norm \cite{alicki2004continuity, audenaert2005continuity}
we may therefore argue that calculation of the Holevo quantity may also be done using the Gaussian states.
We do not attempt to give a rigorous proof here, but by analogy these observations
 suggest a conjecture that a good approximation to Holevo quantity in the strong estimation regime should be
\begin{equation}
\label{eq:strongestimationholevo}
\chi(\{p(x),\rho(x)^{\otimes k}\}) \approx \frac{1}{2} \log \frac{k}{2 \pi e} +  \int \t{d}x\, p(x) \log\frac{\sqrt{F_Q(\rho_x)}}{p(x)} + o(1).
\end{equation}
Therefore, this intuition suggest that collective measurements offer no additional advantage here, unlike in
the weak estimation regime. Note, that one can derive also an other generalized version of \eqref{eq:InfBar}, important for quantum data compression and communication tasks \cite{Hayashi2010, Yang2016}, however, we do not use it here since we deal with classical communication.

Thinking now of entangled input probes in the strong estimation scenario, one should not expect a relation like above to hold in general.
Some fundamental incompatibilities between QFI based and entropy-communication based
 approaches when entangled input probes are considered where underlined in \citep{hall2012does, hall2013metrology}. The most striking is the example of phase estimation using NOON states. While QFI grows as $k^2$, the information communicated is never larger than a single bit as the output state is restricted to a two-dimensional subspace (which also implies that despite large QFI NOON states are not really useful in practical estimation (note, however, that NOON states are not useful in practical estimation \cite{jarzyna2015true, Hayashi2016}).
 Something relatively general, can nevertheless be said. In case channels $\Lambda_x$ are noisy,
 the QFI at the output for optimally entangled input probes generically scales linearly with $k$ and the quantum enhancement amount to a constant factor improvement \cite{demkowicz2014using}.
 Moreover, as argued in \cite{jarzyna2013matrix, jarzyna2015true} one can achieve almost optimal performance utilizing states where
 $k$  probes are divided into groups of $g$ particles where entanglement is present only among the particles belonging to the same group.
 In such a scenario, one can approximate the input state as a product state of large number of groups, number of which will tend to infinity
 while their size will remain constant when $k\rightarrow \infty$. This makes it possible again to apply the reasoning based on quantum local asymptotic normality, and argue that for this class of states \eqref{eq:strongestimationholevo} holds where
 $F_Q(\rho_x)$ is replaced with $\frac{1}{g}F_Q(\Lambda^{\otimes g}(\rho^{g,\t{in}}))$ representing a performance gain thanks
 to the use of entangled input probes.

 In short. While in the weak estimation/communicaton regime we have observed both the effects of input and output super-additivity
 in the strong-estimation regime there seems to be no gain from the use of collective measurements while one may still observe benefits coming
 from entanglement present in the input states.

\section{Examples}
\label{sec:examples}

In this section we provide two examples of communication for which one can easily analyze
issues of input and output super-additivity using the knowledge of the behavior of the QFI and the REQFI and study the validity of our
approximation by comparing it to rigorous calculations.

\subsection{Qubit dephasing channel}
\begin{figure}
\includegraphics[width=0.75\columnwidth]{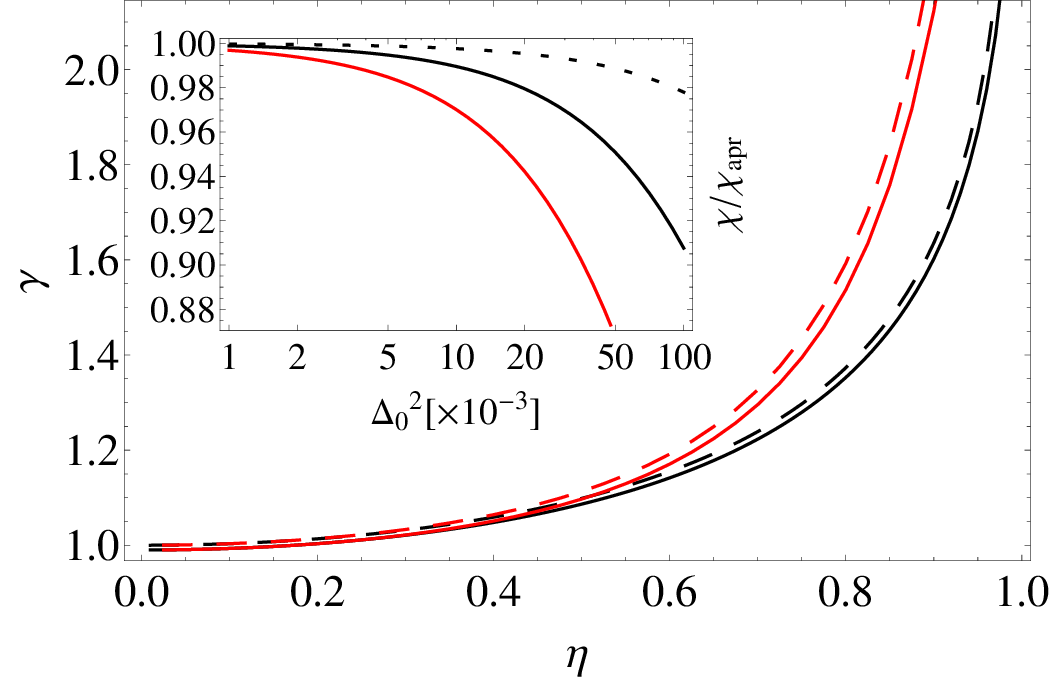}
\caption{
The output super-additive gain factor $\gamma^{(1,\infty)}= C^{(1,\infty)}_{p(\varphi),\Lambda_\varphi}/ C^{(1,1)}_{p(\varphi),\Lambda_\varphi}$
(black) for qubit dephasing channel as a function of dephasing parameter $\eta$  is compared with the case when  entanglement between  two channels inputs is additionally  allowed $\gamma^{(2,\infty)}= C^{(2,\infty)}_{p(\varphi),\Lambda_\varphi}/ C^{(1,1)}_{p(\varphi),\Lambda_\varphi}$ (red); the prior variance is assumed to be $\Delta_0^2=2 \times 10^{-2}$.
The dashed curves represent the same quantities calculated using narrow-prior approximate formulas \eqref{eq:Holevo_approx}.
The inset depicts the validity of our approximation by presenting the ratio of the exact Holevo quantity \eqref{eq:Holevo_qubit} to the approximate expressions  as a function of variance of prior distribution in decoherence-free case $\eta=1$ (black, dotted) as well as in presence od dephasing $\eta=0.9$ for product (black, solid) and  optimally entagled two-channel inputs (red, solid).}
\label{fig:qubit}
\end{figure}
As a first, illustrative example, let us consider a phase encoding qubit channel in presence of partial dephasing:
\begin{equation}
\rho_\varphi = \Lambda_\varphi(\rho^{\t{in}}) = U_\varphi \left( \sum_{i=0}^1 K_i \rho^{\t{in}} K_i^\dagger \right)U_\varphi^\dagger,
\end{equation}
where $U_\varphi=e^{-i\varphi\sigma_z/2}$ is the unitary phase encoding operator, with $\sigma_z$ denoting Pauli $z$ operator,
whereas $K_i$ are Kraus operators of the dephasing map
\begin{equation}\label{eq:Kraus_dephasing}
K_0=\sqrt{\frac{1+\eta}{2}}1\!\!1,\qquad K_1=\sqrt{\frac{1-\eta}{2}}\sigma_z,
\end{equation}
where $\eta$ is the dephasing parameter.

We choose a Gaussian prior probability distribution  $p(\varphi)=\frac{1}{\sqrt{2\pi\Delta_0^2}}e^{-\varphi^2/2\Delta_0^2}$ which is a valid probability distribution on a circle $\varphi\in[-\pi,\pi]$ for small variances $\Delta_0^2$ which is the regime we are interested in.
The Holevo quantity for this model is maximized for input states lying on the equator of the Bloch ball and can be easily calculated yielding
\begin{equation}\label{eq:Holevo_qubit}
C^{(1,\infty)}_{p(\varphi),\Lambda_\varphi} =h_2\left(\frac{1+\eta e^{-\Delta_0^2/2}}{2}\right)-h_2\left(\frac{1+\eta}{2}\right), 
\end{equation}
where $h_2(x)=-x\log x-(1-x)\log(1-x)$ denotes a binary entropy function. In the limit of narrow prior distribution $\Delta_0^2\ll 1$
 the above formula becomes
\begin{equation}\label{eq:Holevo_approx}
C^{(1,\infty)}_{p(\varphi),\Lambda_\varphi}{\approx}\cases{
\frac{\eta\Delta_0^2}{4}\log\frac{1+\eta}{1-\eta} & $\eta<1$\\
-\frac{\Delta_0^2}{4}\log\frac{\Delta_0^2}{4e} & $\eta=1$}.
\end{equation}

Let us now analyze this model with the help of the ideas developed in this paper.
 From \eqref{eq:capacity_individual} we know that $C^{(1,1)}$ can be expressed through the QFI.
 Phase estimation in presence of dephasing is a well understood problem, the QFI is again maximzed for
 states lying on the equator and reads $\max_{\rho^{\textrm{in}}}F_Q[\Lambda_{0}(\rho^{\textrm{in}})]=\eta^2$.
 Hence
 \begin{equation}
 C^{(1,1)}_{p(\varphi),\Lambda_\varphi}\approx \frac{\eta^2\Delta_0^2}{2\ln 2}.
 \end{equation}
 Knowledge of the maximal QFI is also sufficient to calculate approximate $C^{(1,\infty)}_{p(\varphi),\Lambda_\varphi}$ in the decoherence-free case ($\eta=1$) with the help of \eqref{eq:holevopure} and yields the result which agrees with \eqref{eq:Holevo_approx}.  On the other hand, in  presence of decoherence, $\eta<1$, in order to get an approximate expression for $C^{(1,\infty)}_{p(\varphi),\Lambda_\varphi}$
 we need to calculate the REQFI and optimize it over input states. The resulting expression reads $\max_{\rho^{\textrm{in}}}J[\Lambda_0(\rho^{\textrm{in}})]=\frac{\eta}{2}\ln\frac{1+\eta}{1-\eta}$.
 Since the output state $\Lambda_0(\rho^{\textrm{in}})$ is supported on the whole Hilbert space we may utilize
 \eqref{eq:holevomixed} and as a result get the same expression as in \eqref{eq:Holevo_approx}.
 In this case the output super-additivity measure $\gamma^{(1,\infty)}=\frac{1}{2\eta}\ln\frac{1+\eta}{1-\eta}\geq 1$
 proving the generic advantage of the use of collective measurements in this communication protocol.
 Note, however, that for large dephasing $\eta\ll 1$ we have $\gamma^{(1,\infty)}\to 1$ and super-additive behavior of the capacity is lost,
 see Fig.~\ref{fig:qubit}.

To analyze input super-additivity we need to analyze the QFI and REQFI for phase estimation in presence of dephasing when inputs states are allowed to be entangled. For decohrenece free case QFI exhibits Heisenberg limited scaling and we have $F_Q^{(N,\infty)}=N$ which means that with optimal procedure $F_Q^{(N,\infty)}>F_Q^{(1,\infty)}$. Consequently, the optimal QFI for dephasing is super-additive and therefore also capacity for decoherence-free communication exhibits input super-additivity. If, on the other hand, the decoherence is present in our setup, we should consider REQFI instead of QFI. It was shown in \cite{Escher2011, demkowicz2012elusive} that asymptotically, for large number $N$ of two-level particles the QFI per particle is bounded by an expression $F_Q^{(N,\infty)}\leq \frac{\eta^2}{1-\eta^2}$ and the bound is tight in a sense that for large $N$ one can find an entangled input state and measurement for which the inequality is saturated. However, as stated in Sec.~\ref{sec:superadditivity}, REQFI must obey the same classical simulation bound as QFI and since the bound can be saturated by the latter quantity it necessarily also have to be tight for the former one. Therefore we may apply the same reasoning as in the decoherence-free case, this time with a conclusion that  $J^{(N,\infty)}>J^{(1,1)}$. This implies that also in the presence of nonzero dephasing sending entangled input states can improve the capacity
in addition to gains already present thanks to utilizing collective measurements, see figure~\ref{fig:qubit} where the benefits of
 entangling two inputs are depicted and it is clear that $C^{(2,\infty)}_{p(\varphi),\Lambda_\varphi}>C^{(1,\infty)}_{p(\varphi),\Lambda_\varphi}$.

\subsection{Bosonic thermal channel}
\begin{figure}
\includegraphics[width=0.75\columnwidth]{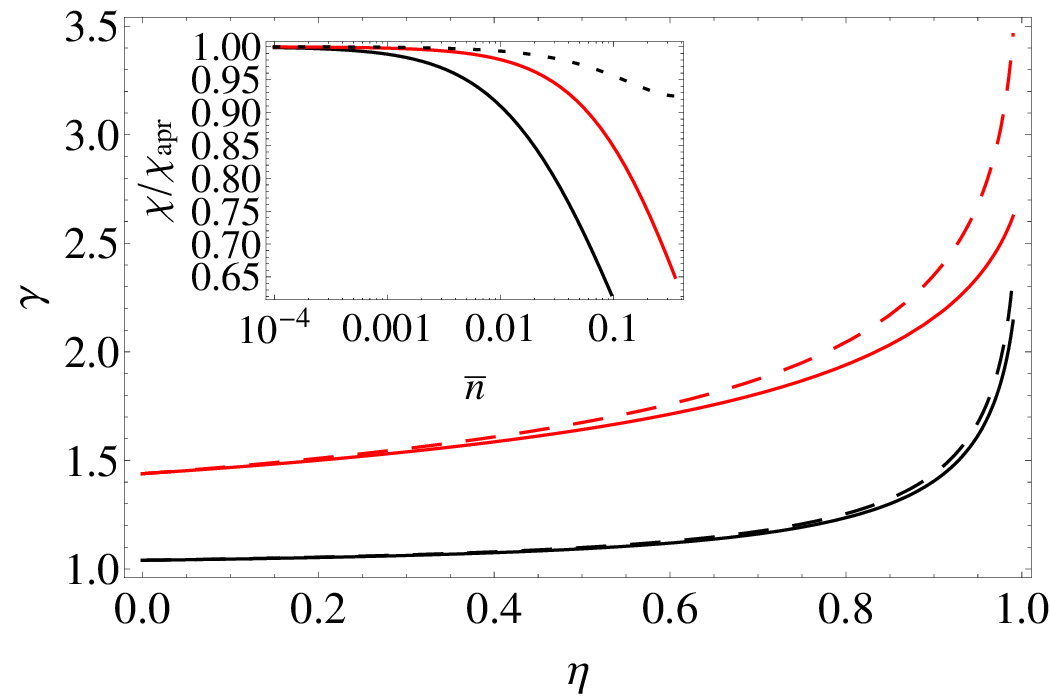}
\caption{The output super-additive gain factor $\gamma^{(1,\infty)}= C^{(1,\infty)}_{p(\alpha),\Lambda_\alpha}/ C^{(1,1)}_{p(\alpha),\Lambda_\alpha}$ for optical communication utilizing coherent states with narrow gaussian amplitude distribution via a thermal lossy channel as a function of transmission coefficient $\eta$ for average input photon number $\bar{n}=0.01$ and thermal number of photons $n_{\t{th}} = 0.1$ (red, solid), $n_{\t{th}}=1$ (black, solid). The dashed lines represent the same quantities calculated using our approximation.
The ratio of the exact Holevo quantity value to the approximated one is depicted in the inset as a function of
the average input photon number which plays the role of the width of input parameter distribution: $n_{\t{th}}=0$, $\eta=0.9$ (black, dotted),
$n_{\t{th}}=0.1$, $\eta=0.5$ (red, solid), $n_{\t{th}}=1$, $\eta=0.99$ (black, solid).  }
\label{fig:bosonic}
\end{figure}

The utility of our approximated formulas applied to the qubit example presented above may be questioned as they are only valid in the
weak estimation regime of narrow prior phase distribution and it is not clear what practical motivation might
 justify the use of such a narrow input phase encoding for communication purposes.
In order to show that the limit is actually of physical interest,
as our second example let us consider a model of optical communication through a thermal channel which is a quantum analogue of classical additive white Gaussian noise channel. As will be clarified below, in this case the weak estimation limit we are interested in appears naturally in practical applications as it corresponds to the regime of small input light intensities---a regime of high relevance and extensively investigated in optical communication literature \cite{Banaszek2012, Guha2011a, Waseda2011, Guha2011}.

Intuitively, the evolution of an input state of light through the thermal channel may be described as mixing a single mode input state with a thermal state $\rho_{\bar{n}_{\textrm{th}}}=\sum_{n=0}^\infty\frac{\bar{n}_{\textrm{th}}^n}{(\bar{n}_{\textrm{th}}+1)^{n+1}}\ket{n}\bra{n}$ with average number of photons $\bar{n}_{\textrm{th}}$ on a beamsplitter with transmissivity $\eta$. Note, that in the extreme case $\bar{n}_{\textrm{th}}=0$ the thermal channel describes pure photon losses. We  assume that the encoding of information is in the amplitude
$\alpha\in\mathbb{R}$ of a coherent state with a Gaussian prior probability distribution $p(\alpha)=\frac{1}{\sqrt{2\pi\bar{n}}}e^{-\alpha^2/2\bar{n}}$, where $\bar{n}$ is the average number of photons per channel use in our communication procedure and simultaneously plays the role of the variance of the amplitude random variable. The encoding procedure is realized by the action of a displacement operator on the input vacuum state $\ket{\alpha}=D(\alpha)\ket{0}$, where $D(\alpha)=e^{\alpha\hat{a}^\dagger-\alpha\hat{a}}$ and $\hat{a},\,\hat{a}^{\dagger}$ are respectively annihilation and creation operators of the input bosonic mode. Since both coherent and thermal states as well as the evolution are Gaussian we may easily express the output state. To do this we apply the methods from \cite{Serafini2005} and get displaced thermal states at the output
\begin{equation}
\rho_\alpha=\Lambda_{\alpha}(\ket{0}\bra{0}) = D(\sqrt{\eta}\alpha)\rho_{(1-\eta)\bar{n}_{\textrm{th}}}D(\sqrt{\eta}\alpha)^\dagger.
\end{equation}
Since the above state is already written in its eigenbasis we can use \eqref{eq:qfiexplicit} and calculate QFI and REQFI for the problem of displacement parameter estimation, which read $F_Q=\frac{4\eta}{1+2(1-\eta)\bar{n}_{\textrm{th}}}$ and $J=2\eta\ln\frac{1+(1-\eta)\bar{n}_{\textrm{th}}}{(1-\eta)\bar{n}_{\textrm{th}}}$ respectively. We can now use these results in order to write communication rate in the regime of small average number of photons $\bar{n}\ll 1$, which is exactly the regime of narrow
prior distributions. First of all, in the absence of thermal environment $\bar{n}_{\textrm{th}}=0$ i.e. for lossy channel, the output states are pure $\rho_\alpha=\ket{\sqrt{\eta}\alpha}\bra{\sqrt{\eta}\alpha}$ so according to \eqref{eq:holevopure} we have
\begin{equation}\label{eq:chi_loss}
\chi\approx \eta\bar{n}\log \frac{e}{\eta\bar{n}},
\end{equation}
whereas even if only a small amount of thermal photons is present the output state is mixed and by \eqref{eq:holevomixed} the rate is reduced to
\begin{equation}\label{eq:chi_thermal}
\chi\approx \eta\bar{n}\log\frac{1+(1-\eta)\bar{n}_{\textrm{th}}}{(1-\eta)\bar{n}_{\textrm{th}}}.
\end{equation}
These expressions agree with the expansion in the average number of photons of the exact Holevo quantity for thermal channel which is given by
\begin{equation}\label{eq:Holevo_bosonic}
\chi=f\left(\sqrt{\beta(2\eta\bar{n}+\beta)}\right)-f\left(\beta\right),
\end{equation}
where the function $f(x)=\left(x+\frac{1}{2}\right)\log\left(x+\frac{1}{2}\right)-\left(x-\frac{1}{2}\right)\log\left(x-\frac{1}{2}\right)$ and $\beta=\frac{1}{2}+(1-\eta)\bar{n}_{\textrm{th}}$ is the diagonal element of the covariance matrix of the thermal state $\rho_{(1-\eta)\bar{n}_{\textrm{th}}}$ \cite{Adesso2006}. The convergence of approximate and exact formulas for small average number of photons can be seen in the inset of figure~\ref{fig:bosonic}.

The above results imply that in the limit of small average number of photons Holevo quantity behavior changes drastically depending whether there are thermal photons in the environment or not. In the first case, according to \eqref{eq:chi_thermal}, the rate scales linearly with the average number of photons. If, however, the environment is in the vacuum state, that is we are dealing with purely lossy channel, we see from \eqref{eq:chi_loss} that rate scales super-linearly with $\bar{n}$. The presence of thermal environment therefore can reduce the rate significantly in the regime of weak signal power.

From the form of the QFI and the REQFI it is also evident that information transmission rates for the considered setup clearly exhibit output super-additivity. In both cases, either the pure lossy channel or thermal channel, we see that Holevo quantity is larger than the respective accessible information, which in the weak estimation limit is clearly visible thanks to the use of formulas \eqref{eq:gammasimple} and
the fact that in our case $J > F_Q$. This super-additive behavior is depicted in figure~\ref{fig:bosonic}.
Note, however, that large thermal noise reduces the gain from using collective measurements and asymptotically for $\bar{n}_{\textrm{th}}\gg 1$ we do not see super-additive behavior $\lim_{n_{\t{th}} \rightarrow \infty}\gamma^{(1,\infty)} = 1$. On the other hand, in the absence of thermal photons, one still gets the advantage from output super-additivity in the regime of small average number of signal photons irrespectively of losses present.

Finally, let us also point out that \eqref{eq:chi_loss} and \eqref{eq:chi_thermal} agree with asymptotic expansion of the capacity of lossy and thermal channels respectively \cite{giovannetti2004classical, Giovannetti2014} in the limit of small average number of signal photons. Therefore based on our approximation we can conclude that in the regime of weak signal power in order to obtain optimal performance of communication it is sufficient to encode information using just the displacement in single quadrature and Gaussian prior probability.
This is an example, where the encoding in the estimation problem considered happens to be the
optimal encoding in the problem of unrestricted capacity optimization, and hence results may be directly related
with the actual channel capacity formulas and not only with the channel ``capacities'' under sub-optimal encodings.


\section{Conclusions}
We have highlighted a connection between communication concepts such as the mutual information and the Holevo quantity on one side and
Fisher information related quantities utilized in quantum estimation theory on the other.  The presented approach allows one to trace the aspects of super-additivity  both at the input as well as at the output stages of the communication protocols provided one operates in the weak estimation regime where the amount of information learned from the measurements is small compared to the prior knowledge. This regime is in particular highly relevant in optical communication utilizing weak light beams.  The main message of the paper is that in this regime the input super-additivity can be linked to the  input super-additivity
 of the QFI $F_Q$ and the REQFI $J$ whereas the output super-additivity is intimately related with the
majorization of $F_Q$ by $J$.
Our results provide also a new operational interpretation for $J$, as it appears naturally in the approximate formula for the Holevo quantity in case of full rank output states
 as well as for $F_Q$ which determines the communication performance in case of pure output states.
Since the symbol encoding in the considered communication protocols were restricted to the ones appearing in the corresponding estimation schemes, the validity of statements on communication super-additivity issues appearing throughout our work is necessarily restricted to these particular encodings. Still, as demonstrated by the example of optical communication in the weak power regime, in some cases simple estimation relevant encodings can be found that lead to the optimal communication performance and hence allow to address the concept of fundamental channel capacity quantity within our approach as well.

\ack
We thank Michal Horodecki for useful discussions. This  work was
supported  by  EU  FP7  IP  project  SIQS  co-financed  by
the Polish Ministry of Science and Higher Education as
well  as  by  the  Polish  Ministry  of  Science  and  Higher
Education  Iuventus  Plus  program  for  years  2015-2017
No.   0088/IP3/2015/73.

\bibliographystyle{iopart-num}
\bibliography{quantum_fisher_and_holevo}

\end{document}